\documentclass[global,twocolumn]{svjour}
\usepackage{latexsym}
\usepackage{graphicx}
\journalname{myjournal}
\begin{document}
\titlerunning{Neutron Halo Isomers and Brilliant Neutron Sources}
\title{Neutron Halo Isomers in Stable Nuclei
and their Possible Application for the Production of
Low Energy, Pulsed, Polarized Neutron Beams of High Intensity
and High Brilliance }
\author{D.~Habs\inst{1}, M.~Gross\inst{1}, P.G.~Thirolf\inst{1}
and P.~B\"oni\inst{2}
}                     
\offprints{}          
\institute{$^1$ Fakult\"at f\"ur Physik,
            Ludwig Maximilians Universit\"at, M\"unchen, D-85748 (Germany)\\
           $^2$ Physik-Department E21, Technische Universit\"at M\"unchen,
D-85748 Garching (Germany)}
\date{Received: date / Revised version: date}
%
\maketitle
\begin{abstract}
We propose to search for neutron halo isomers populated via
$\gamma$-capture in stable nuclei with mass numbers of about
A=140-180 or A=40-60, where the $4s_{1/2}$ or $3s_{1/2}$ neutron
shell model state reaches zero binding energy. These halo nuclei
can be produced for the first time with new $\gamma$-beams
of high intensity and small band width ($\le$ 0.1\%) achievable
via Compton back-scattering off brilliant electron beams, thus
offering a promising perspective to selectively populate these
isomers with small separation energies of 1 eV to a few keV.
Similar to single-neutron halo states for very light, extremely
neutron-rich, radioactive nuclei
\cite{hansen95,tanihata96,aumann00}, the low neutron separation
energy and short-range nuclear force allow the neutron to tunnel
far out into free space much beyond the nuclear core radius. This
results in prolonged half-lives of the isomers for the
$\gamma$-decay back to the ground state in the 100 ps-$\mu$s
range. Similar to the treatment of photodisintegration of the
deuteron, the neutron release from the neutron halo isomer via a
second, low-energy, intense photon beam has a known much larger
cross section with a typical energy threshold behavior. In the
second step, the neutrons can be released as a low-energy, pulsed,
polarized neutron beam of high intensity and high brilliance,
possibly being much superior to presently existing beams from
reactors or spallation neutron sources.
\end{abstract}

\section{Introduction}

Presently, thermal and cold neutron beams are produced at large-scale
facilities like reactors or spallation sources via moderation of MeV
neutrons down to the meV regime. Moderators and shielding result in very large
sources with $\sim$10~m diameter and accordingly reduced flux density.
  We want to realize a new intense, brilliant neutron source by
first populating weakly bound, neutron halo isomers which, after
stopping, are used to release a directed, low-energy neutron beam
when irradiating them with a second laser or an X-ray beam.

\begin{figure*}[t!]
\centerline{\includegraphics[width=.8\textwidth]{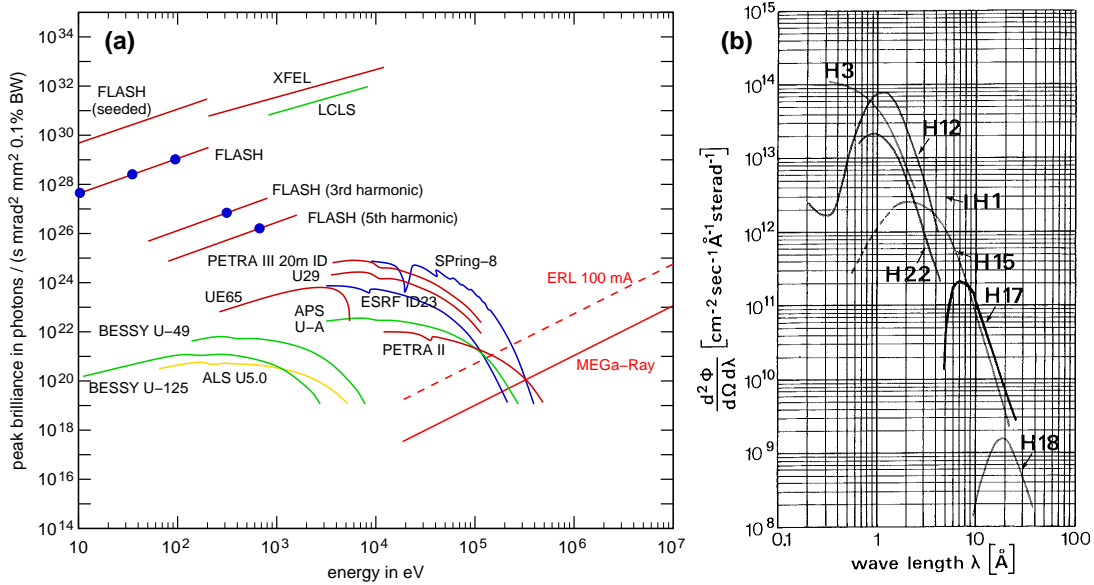}}
   \caption{ (a) Brilliance of X-ray and $\gamma$-ray sources
               \cite{XFEL02,barty10} compared
   with (b) the brilliance of neutron beams at the ILL reactor
            (Grenoble) comprised in the Yellow Book~\cite{ILL88}
          (BW=Band Width).}
   \label{fig1}
\end{figure*}

Let us first compare the brilliances of X-ray beams, neutron beams, 
and $\gamma$ beams, which are all charge neutral.
12.4 keV X-rays and 81.8 meV neutrons are penetrating particles 
with a wavelength of 1 {\AA}  that is matched to typical atomic
distances in solids \cite{hercules94}. A significant advantage of
neutrons compared to X-rays is their interaction with nuclei and their
magnetic moment. Hence, neutrons are deeply penetrating and can be
used for the investigation of structural and magnetic volume properties of
materials on a microscopic scale. In particular, neutrons interact also
strongly with light elements, a property that is important for
investigating soft matter and biological materials. The comparison
of the XFEL \cite{XFEL02} and the European Spallation Source ESS
\cite{ESS09} in view of the decision to install probably both large-scale
facilities in Europe indicates the complementarity between
X-rays and neutron beams, although the brilliance of
X-FELs ($10^{32}$/(mm$^2$ mrad$^2$ s (0.1\%BW)) is some 30 orders of
magnitude higher than that of the most intense neutron beams at the ILL
reactor \cite{ILL88}, which provides neutron beams with
$10^2-10^3$ /(mm$^2$ mrad$^2$ s (0.1\%BW)). The brilliance of various
X-ray sources and beam lines at the ILL reactor are shown in Fig.~\ref{fig1}.

\begin{figure}[b]
\centerline{\includegraphics[width=.47\textwidth]{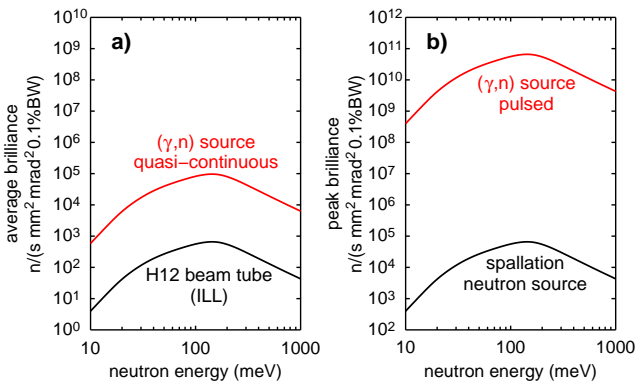}}
   \caption{(a) Average brilliance of continuous neutron sources
   and (b) peak brilliance of pulsed neutron sources as a function of neutron
energy. For the reactor-based source
we converted the curve of the H12 beam line of the
ILL given in Fig.~\ref{fig1}. For the peak brilliance of spallation sources we
increased the maximum average brilliance of a reactor by a factor of 100.
Only the rough energy dependence of the brilliances are relevant, since details
depend on the reflectivities of neutron guides. Also the expected brilliances
for the new $(\gamma,n)$ sources using halo isomers are shown in red,
where the curves represent rough order of magnitude estimates.}
   \label{fig1a}
\end{figure}

For comparison, in Fig.~\ref{fig1a} we have converted the neutron
values of the most brilliant beam line H12 of the ILL reactor
(Fig.~\ref{fig1}b) to the standard representation used for X-
and $\gamma$-rays, changing the neutron wavelength given in {\AA}
~by the nonlinear dispersion to neutron energies given in meV by

\begin{equation}
   (\lambda_n[\mbox{\AA}])^2=\frac{81.81}{E_n[\mbox{meV}]}.
\end{equation}

The brilliances shown in Fig.~\ref{fig1a} for the new neutron
source envisaged for the ELI-Nuclear Physics project
\cite{ELI-NP10} using halo isomers and $\gamma$-beams will be
explained later. The comparison of the different neutron
brilliances points to a major breakthrough in neutron physics if
the two-step production process via neutron halo states proves
successful.

Currently, thermal and cold neutron beams are extracted as
continuous beams from nuclear reactors or as pulsed beams from
spallation sources. In both cases the primarily produced neutrons
have an energy of several MeV. Therefore, to make them useful for
experiments in solid-state physics, they are moderated to thermal
energies by elastic scattering in heavy water (thermal neutrons),
liquid hydrogen or deuterium (cold neutrons), or other moderators.

A typical moderator of a nuclear research reactor like the FRM II
in Garching or the ILL in Grenoble consists of a tank with a
radius of approximately 1 m of heavy water, where the very small
absorption cross section for neutrons and the similar masses of
the neutrons and the moderator are decisive for an optimum
performance of the moderation process. This tank is surrounded by
a second tank of light water for shielding and cooling purposes.
The neutron intensity of a high-flux reactor for neutron research
is about $10^{18}$n/s \cite{hercules94} corresponding to an
optical light source or a candle of 200 mW, if one replaces the
neutrons by photons -- illustrating that the presently available
intensity of neutrons is very small when compared with an X-ray
tube or a laser.

The neutrons are then extracted by beam tubes pointing towards the
moderator followed by neutron guides. Including the biological shielding,
the neutrons can be used after traversing a distance of
approximately 5 m, i.e. they exit through a surface of $\pi\cdot
10^6$ cm$^2$, resulting in a typical overall flux $F_{reac} \simeq 3\cdot
10^{11} $n/(cm$^2$s). The neutron moderators typically emit radiation
into a solid angle of $\approx$1 sterad. If one takes this solid angle
into account, this compares well with the flux of beam line H12
at the ILL given in the Yellow Book \cite{ILL88} by
$F_{H12} = 2.9 \cdot 10^{10}$ n/(cm$^2$s).

At spallation sources, the liquid mercury target for neutron
production is hit by a pulsed proton beam with a repetition rate
of typically 50 Hz and 1 GeV energy. It is expected that future spallation 
sources will reach up to 5 MW target power, which is about an order of
magnitude smaller than the thermal power of larger research
reactors. The average target design flux of the ESS is $3\cdot
10^{14}$n/(cm$^2$s), which is a factor of 5 smaller than, e.g., the
flux of the beam line V4 at the ILL reactor with $1.5\cdot
10^{15}$n/(cm$^2$s). Since the geometrical extraction efficiency
of a spallation source \cite{koester10} is somewhat better than
that of a reactor, the average extracted neutron flux may be only a
factor of 2 lower than that of a reactor. However, the peak
brilliance is about a factor of 100 higher \cite{ESS09,SNS07}.
Spallation sources as well as reactors produce large amounts of
radioactivity due to the high-energy neutrons and the fission and
fragmentation products. Moreover, they require large amounts of
shielding, safety and security precautions. Therefore, the direct
production of neutrons using $\gamma$-rays is a very attractive
rather low-cost solution for the investigation of materials, when
small and highly brilliant beams are required.

\begin{figure}[t]
\centerline{\includegraphics[width=.47\textwidth]{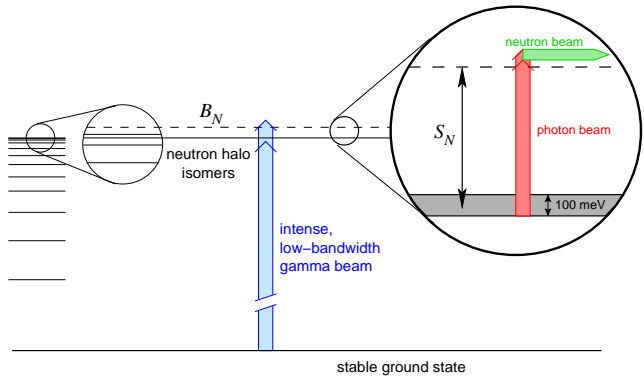}}
   \caption{Schematic picture of the new neutron production scheme:
   With the $\gamma$-beam exciting the neutron halo isomers of neutron
  separation energy $S_N$ below the binding energy $E_B$. The left level scheme
  shows the increasing number of compound nuclear reasonances with excitation
energy. The halo isomer is admixed to several high-lying resonances, resulting
in halo isomers with different binding energies.    
  The two blue arrows indicate the width of the $\gamma$ beam.
   Then in a second step, a photon beam of much lower energy, shown in red,
   generates the neutron beam by dissociating the neutron halo state.}
   \label{fig3}
\end{figure}

In Fig.~\ref{fig3} we show schematically the new neutron beam production
scheme, where we want to produce a brilliant pulsed neutron beam directly 
without moderation by exciting in a first step the neutron halo isomers
at excitation energies just below the neutron separation threshold
via a $\gamma$-capture reaction using brilliant $\gamma$-beams of 6-8 MeV.
A neutron halo isomer is a longer-lived nuclear state, where one neutron
of the nucleus is excited into a very weakly bound configuration, being
long-lived and extending into regions far away from the nuclear core.
 The narrow band width of the $\gamma$-beam is
essential for a selective production of the isomers. The $(\gamma$,n)
reaction above threshold is suppressed due to a phase space factor
in neutron production. In a second step, the neutrons of these halo
isomers are effectively released by a second intense laser or photon beam
of narrow band width due to the  large cross section, producing a
brilliant neutron beam.

\begin{figure}[t]
\centerline{\includegraphics[width=.47\textwidth]{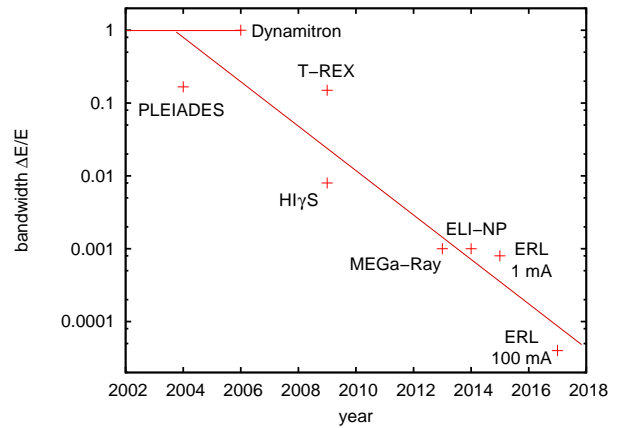}}
   \caption{Band width of high-energy $\gamma$-beams ($\approx$ 10 MeV)
as a function of time.}
   \label{fig4}
\end{figure}

\begin{figure}[t]
\centerline{\includegraphics[width=.47\textwidth]{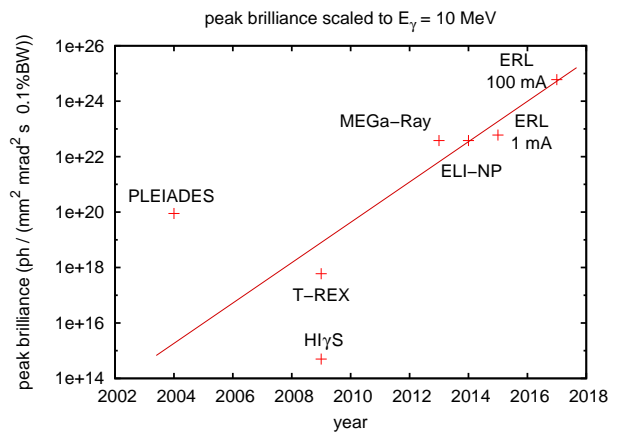}}
   \caption{Peak Brilliance of high-energy $\gamma$-beams ($\approx 10$ MeV)
            as a function of time.}
   \label{fig5}
\end{figure}

Such a concept only became feasible due to the development of
very brilliant $\gamma$-beams, where the $\gamma$-rays are
produced by incoherent Compton back-scattering of laser light from
brilliant high-energy electron bunches. Fig.~\ref{fig4} and
Fig.~\ref{fig5} show the rapid progress of $\gamma$-beams with
respect to band width (Fig.~\ref{fig4}) and peak brilliance
(Fig.~\ref{fig5}) with time, starting with the bremsstrahlung
spectrum of the Stuttgart Dynamitron \cite{kneissl06}, which still
had a very large band width, and proceeding also to projected
characteristics of future facilities (see below).

For Compton back-scattering in a head-on collision \cite{hartmann05}
the $\gamma$-energy is given by:

\begin{equation}
   E_{\gamma}=\frac{4\gamma_e^2 E_L}{1+
              (\gamma_e \Theta_{\gamma})^2 + 4\gamma_e E_L/mc^2}
\end{equation}

where the $\gamma_e$-factor denotes the energy of the
electron beam $E_e=\gamma_e\cdot mc^2$, $\gamma$-energy is the 
energy of the $\gamma$-photon, $\Theta_{\gamma}$ is the angle 
between $\gamma$-beam and the the e-beam, and $E_{L}$ is the laser 
photon energy. The energy $E_{\gamma}$ decreases with increasing
$\Theta_{\gamma}$. A small band width of the $\gamma$-beam
requires a small energy spread of the electron bunches $(\Delta
\gamma_e/\gamma_e)$, a small band width of the laser energy
$(\Delta E_L/E_L)$ , a very good emittance of the electron beam
with a small opening and a small opening angle of the laser beam.

At the HI$\gamma$S facility (Duke University, USA) the primary
photons are produced by an FEL using undulators and electrons from a
storage ring. Then in a second step these FEL-photons are back-scattered
from the circulating electron beam \cite{weller09}.
C.~Barty and his group at the Lawrence Livermore
National Laboratory (LLNL) developed already three generations
of incoherent Compton back-scattering sources: PLEIADES \cite{pleiades04},
T-REX \cite{trex10} and MEGa-Ray \cite{barty10}, each based on a ``warm''
electron linac and a fibre laser for back-scattering. However, the electron
linac technology was switched from S-band technology (4 GHz) for T-REX to
 X-band technology (12 GHz) for MEGa-Ray. The  MEGa-Ray
$\gamma$-beam runs with a macro pulse structure
of 120 Hz using  0.5 J, 2 ps laser pulses, which are recirculated 100 times
with 2 ns bunch spacing in a ring-down cavity. A similar $\gamma$-facility is
planned for
the ELI-Nuclear Physics project (ELI-NP) in Romania \cite{ELI-NP10},
also based on a``warm'' linac like that used at MEGa-Ray, however,
designed for $\gamma$-energies up to 19 MeV. At JAERI in Japan,
R. Hajima and coworkers are developing  a Compton back-scattering
$\gamma$-beam using an energy recovery linac (ERL) and superconducting ``cold''
cavities \cite{hajima09}. For smaller electron bunch charges (10 pC),
a very low normalized emittance of 0.1 mm mrad can be obtained
from the electron gun. For the back-scattered laser
light a high-finesse enhancement cavity is used for recirculating the photons.
The quality of the electron beam from the ERL can be preserved by running
with higher repetition rate. Switching from a 1 mA electron current to
a 100 mA current, the peak brilliance and band width can be improved
significantly \cite{hajima10,ERL-GAMMA08}. Fig.~\ref{fig4} and Fig.~\ref{fig5}
show the rapid improvement of the $\gamma$ beams, which will result in
corresponding improvements of the neutron beams (see section 3).

A one-step neutron production process,
i.e. releasing the neutrons in a $(\gamma,n)$ reaction just
above the neutron binding energy is rather inefficient because the nucleus
acquires from the high-energy $\gamma$ quantum a recoil energy $E_{rec}$
according to

\begin{equation}
    E_{rec}/A= \frac{E_{\gamma}^2}{2\cdot Mc^2\cdot A^2} .
\end{equation}

For  mass number A=180 and $E_{\gamma}=$ 7 MeV, we
obtain $E_{rec}/A$ = 500 meV.
Thus also the emitted neutron acquires a recoil energy in $\gamma$ direction
of 500 meV,
which is rather large compared to the 25 meV of thermal neutrons.
Only for a very small emission opening angle, small neutron
energies could be obtained, where the $\gamma$ recoil momentum is compensated.
Using the neutron halo isomers, which are stopped in the target within
a few ps, the $\gamma$-recoil problem can be avoided.

In the following we first (section 2) describe the expected properties of
neutron halo states, their production, dissociation, and experiments
for a first direct observation of neutron halo isomers in stable nuclei.
In section 3 we discuss the properties of the new neutron beams like
flux and brilliance. In section 4 we show possible experiments with the
new neutron beam.

\section{Neutron Halo States}
\label{halo}

The concept of a neutron halo nucleus exists for more than 50
years. A weakly bound neutron can tunnel out of the nuclear
potential with the short-range forces into a large volume of free
space. Due to the small binding energy the wave function by the
uncertainty principle can reach out far into classically forbidden
regions. A halo state is quite different from an atomic Rydberg
state, where long-range Coulomb forces allow for an extended wave
function of an electron with weak binding in classically allowed
regions. In nuclear physics the front runner of the halo nucleus
concept was the deuteron.
 The spin S=1 deuteron ground state with a neutron separation energy
 $S_n$=2.2246 MeV has a halo radius a = 4.3 fm which is much
larger than the radius of R=2.0 fm for the triplet nuclear
potential between neutron and proton with a depth $V_t$ = -36 MeV
\cite{hansen95,segre77}. Thus the nucleons in the deuteron spend
only one third of the time within the range of the nuclear forces
and two thirds of the time they are at distances larger than R.

\subsection{Light, Extremely Neutron-rich Radioactive Neutron Halo Nuclei}

Starting in 1983 with $^{11}$Be \cite{millener83} and $^{11}$Li
\cite{tanihata85,hansen87}, a whole region of light, extremely neutron-rich
halo nuclei was studied, establishing halos for, e.g. , $^6$He,$^{11}$Li,
$^{11}$Be, $^{14}$Be, and $^{17}$B 
\cite{riisager94,hansen95,tanihata96,aumann00}.
One of the best examples for a neutron halo nucleus is $^{11}$Be, where
the ground state 2s$_{1/2}^+$ has a neutron separation energy
$S_n$=0.50 MeV and a radius of 6.0 fm, while the first and only excited
$1p_{1/2}^-$ state
has a neutron separation energy $S_n$=0.18 MeV and a radius of 5.7 fm.
These radii are large compared to the radius of the nuclear core of R=2.5 fm
\cite{hansen95,aumann00}. These halo states are predominantly single particle
states, where most physics is dominated by the external asymptotic
wave function.

\begin{figure}[t]
\centerline{\includegraphics[width=.47\textwidth]{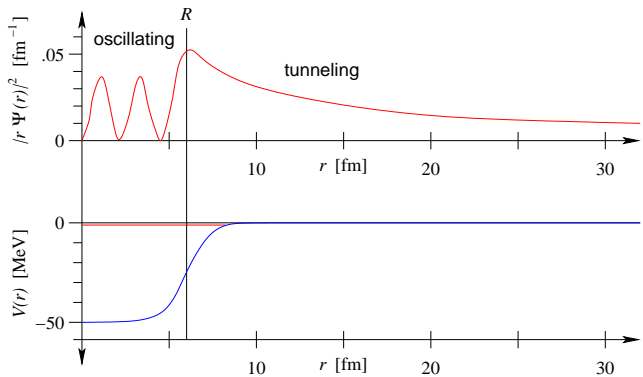}}
   \caption{Square of wave function times $r^2$ for Woods-Saxon potential
           of neutron halo state with N=4 nodes and angular momentum l=0. 
          The lower part shows the potential (blue) with the weakly bound state
           (red).}
   \label{fig6}
\end{figure}

The wave function of the neutron outside the core nucleus  $r\ge R$
is given by

\begin{equation}
\Psi (r)= \left(\frac{\kappa}{2 \pi}\right)^{1/2}\left(\frac{e^{-\kappa r}}{r}\right)
\end{equation}

where the parameter $\kappa$, which has the dimension of a reciprocal
length, is given by

\begin{equation}
\kappa=\sqrt{2\mu S_n}/\hbar=1/a.
\end{equation}

The quantity $S_n$ is the neutron separation energy, 
$\mu$ is the reduced mass and $a$ is the scattering length. 
Inside a square-well potential of depth $V$ ($r\le R$)
the wave function of the neutron is

\begin{equation}
\Psi (r)=\left(\frac{\kappa}{2\pi}\right)^{1/2}
\left(\frac{k^2+\kappa^2}{k^2}\right)^{1/2}
\left(\frac{\sin(kr)}{r}\right)
\end{equation}

with

\begin{equation}
k=\sqrt{2\mu (V-S_n)}/\hbar.
\end{equation}

For the large excitation energy $(V-S_n)$ the wave function has several
nodes. For normalization of the wave function we make the approximation
that the outer integral ($r\ge R$) is much larger than the inner one.
For a given potential we obtain an eigenvalue $S_n$.
In the top part of Fig.~\ref{fig6} we show the squared wave function
(multiplied by $r^2$) of a neutron halo state with N=4 nodes and
angular momentum l=0, calculated for the Woods-Saxon potential indicated
in the bottom part of Fig.~\ref{fig6}.

\begin{figure}[h]
\centerline{\includegraphics[width=.5\textwidth]{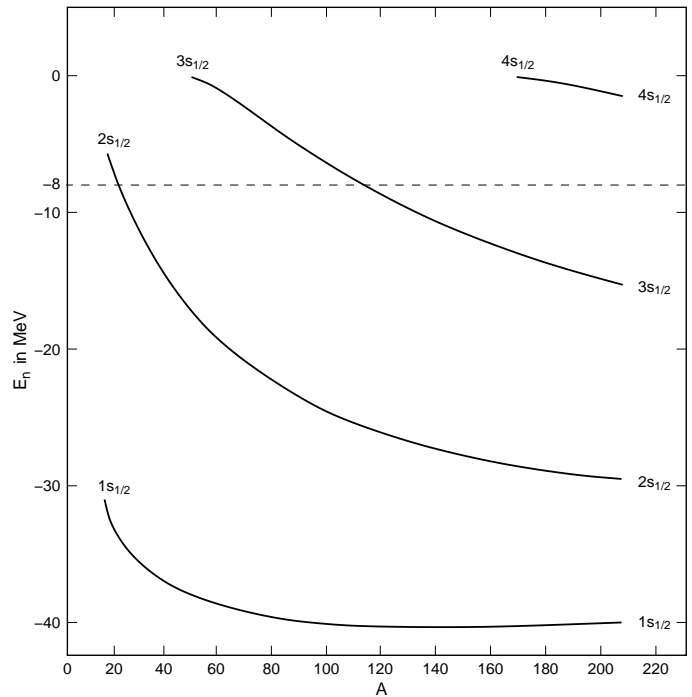}}
   \caption{ Energies of neutron orbits with angular momentum l=0, calculated
    for a Woods-Saxon potential and a spin orbit $l\cdot s$ potential.}
   \label{fig7}
\end{figure}

The halo states of these extremely neutron-rich, radioactive nuclei
cannot be used for a neutron source because the neutron separation energies
are still a few 100 keV large and the target nuclei decay by very fast
$\beta$ decay.

\begin{figure*}[t]
\centerline{\includegraphics[width=0.8\textwidth]{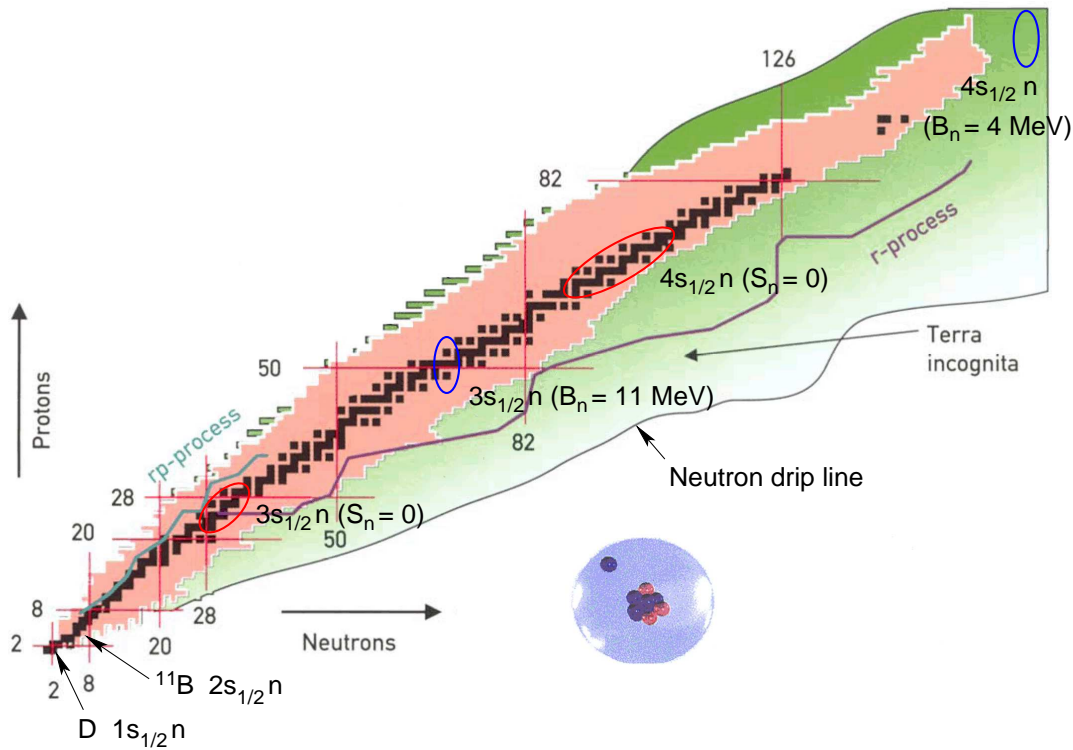}}
   \caption{Nuclidic chart with stable nuclei shown in black. The orange
area contains nuclei, where some properties have been measured,
while the green area called ``terra incognita'' covers unknown
nuclei up to the neutron and proton drip line. Also indicated are
the deuteron D, the region of light, neutron-rich halo nuclei and
the predicted new regions of halo nuclei (red ellipse). The
excitation energies are close to the neutron binding energy, where
for A=(50-60) 3s$_{1/2}$ halo neutrons and for A=(150-180) the
4s$_{1/2}$ halo neutrons should lead to isomers. Also shown as
blue ellipses are the regions of the 3s$_{1/2}$ and 4s$_{1/2}$ for
ground state binding energies. Schematically also a 1-neutron halo
nucleus with the dense core of protons and neutrons is shown at
the lower right.}
 \label{fig8}
\end{figure*}

\subsection{Neutron Halo Isomers in Stable Nuclei}

Within our concept for the production of novel neutron beams we
first want to search for two new islands of halo states for stable
nuclei just below the neutron binding energy $S_n$. In these
nuclei with A $\approx 140-180$ or A $\approx 40-60$, the
$4s_{1/2}$, respectively $3s_{1/2}$, the neutron halo state just
approaches zero binding energy with a very shallow A dependence
\cite{bohrmottelson69}. This dependence can be seen in
Fig.~\ref{fig7}. We are only interested in l=0 neutron orbits,
because only for these the centrifugal barrier vanishes. In
neutron orbit calculation a central spherical Woods-Saxon
potential with

\begin{eqnarray}
V(r)=-V\cdot f(r)\nonumber\\
f(r)=[1+\exp{((r-R)/a)}]^{-1}\\
-V=\left(-51 +33\frac{N-Z}{a}\right) MeV \nonumber
\end{eqnarray}

with R=$r_0A^{1/3}$ and $r_0$=1.25 fm, a=0.65 fm is used.
Also a spin orbit ${\bf l\cdot s}$ potential was included
\cite{bohrmottelson69}. Due to the difference
in the centrifugal barrier, l=0 orbitals are preferred to obtain more
binding for loosely bound neutrons.

In Ref. \cite{bohrmottelson69} also the experimental s-wave
strength function measured with slow neutrons is shown as a
function of mass number A. This strength function is defined as
the ratio of the reduced neutron width $\Gamma_n^{(0)}$ divided by
the average level spacing D and is a measure for the total amount
of one-particle contribution per energy interval in the spectrum
of resonances. One finds an accumulation of s-wave strength in the
region A=140-180 and A=50-60. Since $1/\kappa=a$, very extended
weakly bound systems have been observed in thermal neutron
scattering looking at the scattering length \cite{sears92}. They
are compiled in Table 1, where a maximum scattering length of +49
fm corresponds to a very extended bound halo state, and was
observed indirectly by neutron scattering.

\begin{table}[h]
\caption{Large scattering lengths $a$  \cite{sears92} and neutron
binding energies $S_n$ \cite{audi03} for some selected nuclei with
A=140-180 and with A=40-60.}
\bigskip
\begin{center}
\begin{tabular}{ccc} \hline
isotope & $a$~[fm] & $S_n$~[MeV]\\ \hline
$^{143}$Nd & 14.2 & 6.1 \\
$^{145}$Nd & 14.2 & 5.8 \\
$^{147}$Pm & 12.6 & 7.6 \\
$^{147}$Sm & 14.  & 6.3 \\
$^{150}$Sm & 14.  & 8.0 \\
$^{164}$Dy & 49.2 & 7.7 \\
$^{166}$Er & 10.5 & 8.5 \\
$^{174}$Yb & 19.3 & 7.5 \\
$^{174}$Hf & 10.9 & 8.5 \\
$^{180}$Hf & 13.2 & 7.4 \\ \hline
$^{45}$Sc  & 12.3 & 11.3\\
$^{58}$Fe  & 15.  & 10.0\\
$^{58}$Ni  & 14.4 & 12.2\\ \hline
\end{tabular}
\end{center}
\end{table}

Since low-energy neutron scattering and  $\gamma$-capture reactions usually
populate different spin and parity states, we can only predict the existence
of such weakly bound neutron states in general, but nothing has been measured
until now by $\gamma$-capture reactions. Thus we want to explore these states
by irradiating stable nuclei with mass number A= 140-180 and A=50-60 with
$\gamma$-beams. Since the small binding energy $S_n$ of the isomers depends
critically on many details like nuclear
deformation, their detailed energies can only be explored experimentally.
On the other hand, only the new $\gamma$-beams with very small band width
allow to populate these states with sufficient selectivity to allow for
their detection.

We also show these new regions of halo isomers in the nuclidic
chart of Fig.~\ref{fig8} as red ellipses. 

While for the ground
state configuration one expects the $3s_{1/2}$ neutron orbit close
to the magic neutron number N=82 and A$\approx 138$ and the
$4s_{1/2}$ neutron orbit 20 neutrons below the magic neutron
number N=184 and a corresponding A$\approx 270$ (blue ellipses of
Fig.~\ref{fig8}), these $3s_{1/2}$ and $4s_{1/2}$ neutron regions
for the halo isomers with close to zero binding energies occur in
very different regions. The A $\approx 50$ ground state nuclei
have a neutron binding energy of about 11 MeV, the A$\approx$ 160
nuclei of about 8.6 MeV. One can now understand the shift of the
neutron orbital energy with mass number A of Fig.~\ref{fig7} due
the the A-dependence of the Woods-Saxon potential. The radius R
increases with A and also the potential -V  becomes more negative
with A, explaining this shift when solving for the eigenvalues of
the $3s_{1/2}$ and $4s_{1/2}$ states. Thus while usually the
$4s_{1/2}$ ground state occurs for inaccessible super-heavy nuclei,
we can study their orbits for A=140-160 nuclei at binding energies
close to zero. We can test the method for the $3s_{1/2}$ orbit
comparing the A$\approx$50 and A$\approx$138 regions. For odd
neutron states close to the magic number N=82 we can nicely
identify the $3s_{1/2}$ orbital.

The $4s_{1/2}$ or $3s_{1/2}$ states must combine with the proper
spin and parity state of the core (A-1) nucleus to be excited with
an E1 transition from the spin and parity of the ground state of
the nucleus with mass A. Thus the neutron halo state is frequently embedded
in the continuum with respect to the ground state of the (A-1)
nucleus, because some excited (A-1) core states with known energy
may be required to fulfill the spin and parity requirements (see
Table 2 and Table 3).

Let us now discuss the damping between a compound nucleus resonance and
a neutron halo state. We know the solution  for the situation,
when the $4s_{1/2}$ state sits in the middle of the center of the
excited compound nucleus resonance. Due to the very small
energy difference in a two-state model including perturbation theory,
a strong mixing occurs. We can consider the
neutron as a classical particle oscillating in the deep potential of about
50 MeV above the bottom of the potential. However, close to the border
of the Woods-Saxon potential the neutron becomes slow and needs time
to explore the tunneling region. For a halo state with 1 eV binding
energy we obtain a radius $1/\kappa \approx$ 4000 fm, which is extending
very far out of the core by tunneling. An increase of the binding energy by
100 meV compared to the depth of 50 MeV and an oscillator energy
$\hbar \omega_0$=7 MeV corresponds to a change by $\approx 10^{-8}$.
Thus the wave function change inside the potential is
very small and minor fluctuations in the shape of the potential
e.g. of the radius can still lead to an eigenvalue at the changed energy.
Thus the neutron will move a little slower inside, but will tunnel
out a little less. Thus a neutron injected with a 100 meV lower kinetic
energy will have a negligible change in its oscillatory dynamics.
Due to Pauli blocking, scattering in the central core is suppressed.

\begin{table*}[t!]
\caption{Table of halo isomers with stable ground state for A=140-160.
The column ``\%'' lists the abundance of the particular isotope in \%,
$I_g$ is the ground state spin, $I_{core}$ the spin of the (A-1) core with
the known excitation energy $E_{core}$ of the excited core,
$S_n$ are the tabulated neutron binding energies from Ref.\cite{audi03}.}
\bigskip
\begin{center}
\begin{tabular}{ccccccccccc} \hline
A  &Z & \%   & $I_g$ &$I_{core}$&$I_{core-g}$&$E_{core}$ & $S_n$ &$E_{tot}$&$I_{i}$&$I_{i0}$ \\
  &  &      &       &        &         & [keV]     &[keV]  & [keV]&&\\ \hline
142&Ce&  11.1&$0^+$  &$3/2^-$&$7/2^-$   & 662.06    &7169.7 &7831.76  &$1^-$ &$3^-$\\
142&Ce&  11.1&$0^+$  &$1/2^-$&$7/2^-$   & 1137.0    &7169.7 & 8307.   &$1^-$ &$4^-$\\
143&Nd&  12.2&$7/2^-$&$4^+$  &$0^+$    & 2101.32    &6123.57& 8224.89 &$5/2^+$&$1/2^+$ \\
144&Nd&  23.8&$0^+$  &$3/2^-$&$7/2^-$  &742.02     &7817.03& 8559.    &$1^-$  &$3^-$ \\
144&Nd&  23.8&$0^+$  &$1/2^-$&$7/2^-$  &1305.82    &7817.03& 9123.    &$1^-$&$4^-$ \\
145&Nd&   8.3&$7/2^-$&$4^+$  &$0^+$    &1314.54    &5755.29& 7070.    &$5/2^+$&$1/2^+$ \\
146&Nd&  17.2&$0^+$  &$3/2^-$&$7/2^-$  &67.22      &7565.23& 7632.45  &$1^-$  &$3^-$ \\
148&Nd&   5.7&$0^+$  &$1/2^-$&$5/2^-$   &214.59     &7332.8 & 7547.4  &$1^-$  &$3^-$ \\
148&Nd&   5.7&$0^+$  &$3/2^-$&$5/2^-$   &324.67     &7332.8 & 7657.5  &$1^-$  &$2^-$ \\
149&Sm&  13.6&$7/2^-$&$4^+$  &$0^+$    &1180.26    &5871.1 & 7051.4   &$5/2^+$&$1/2^+$ \\
150&Nd&   5.6&$0^+$  &$1/2^-$&$5/2^-$  &165.09     &7380.1 & 7545.2   &$1^-$  &$3^-$ \\
150&Sm&   7.4&$0^+$  &$3/2^-$&$7/2^-$  &350.04     &7986.7 & 8336.7   &$1^-$  &$3^-$ \\
151&Eu&  47.9&$5/2^+$&$2^-$  &$5^-$    & 70.       &7933.  & 8003.    &$3/2^-$&$7/2^-$ \\
152&Sm&  26.6&$0^+$  &$3/2^-$&$5/2^-$  &  4.82     &8257.6 & 8262.4   &$1^-$  &$2^-$ \\
153&Eu&  52.1&$5/2^-$&$3^+$  &$3^-$    & 113.97    &8550.29& 8664.26  &$3/2^-$&$3/2^+$ \\
154&Sm&  22.6&$0^+$  &$3/2^-$&$3/2^+$  &  35.84    &7967.1 & 8002.9   &$1^-$  &$1^+$ \\
154&Gd&   2.1&$0^+$  &$3/2^-$&$3/2^-$  &   0.0     &8894.71& 8894.71  &$16-$  &$1^-$ \\
155&Gd&  14.8&$3/2^-$&$0^+$  &$0^+$    &   0.0     &6435.22& 6435.22  &$1/2^+$&$1/2^+$ \\
156&Gd&  20.6&$0^+$  &$3/2^-$&$3/2^-$  &   0.0     &8536.39& 8536.39  &$1^-$  &$1^-$ \\
156&Dy&   0.1&$0^+$  &$3/2^-$&$3/2^-$  &   0.0     &9441.  & 9441.    &$1^-$  &$1^-$ \\
157&Gd&  15.7&$3/2^-$&$0^+$  &$0^+$    &   0.0     &6359.89& 6359.89  &$1/2+$ &$1/2^+$ \\
158&Gd&  24.8&$0^+$  &$3/2^-$&$3/2^-$  &   0.0     &7937.39& 7937.39  &$1^-$  &$1^-$ \\
158&Dy&   0.1&$0^+$  &$3/2^-$&$3/2^-$  &   0.0     &9056.  & 9056.    &$1^-$  &$1^-$ \\
159&Tb& 100.0&$3/2^+$&$0^-$  &$3^-$    &  110.3    &8133.1 & 8243.4   &$1/2^-$&$7/2^-$ \\
159&Tb& 100.0&$3/2^+$&$1^-$  &$3^-$    &  115.5    &8133.1 & 8248.6   &$1/2^-$&$5/2^-$ \\  \hline
\end{tabular}
\end{center}
\end{table*}

The damping width of the compound nucleus resonance coupling into the
neutron halo isomer is large compared to the $\gamma$-decay width
inside the nuclear potential (the resonance, via the strong interaction, can
easily eject a quasi-free neutron), but the damping width of the
halo isomer coupling back into the compound nucleus resonance is very weak,
because the spatial probability for the neutron to reach an overlap
with the core region again is very small. The short range nuclear
interaction inside the core results in a strong coupling and fast ejection
of the neutron. In the halo state the neutron is slow and with a very high
probability, in free space without interaction and couples back weakly.
This difference in damping and decay widths is characteristic for halo states.
Usually the neutron halo state will be fragmented into many components,
however, here only the few strong components are of interest, because they have
several orders of magnitude higher E1 excitation strength, due to the 
large dipole moment of the halo state. 

On the one hand, the long lifetime of the halo isomer implies a narrow width
and a weak coupling to the $\approx$100 meV \cite{segre77} broad compound
nucleus resonance. On the other hand, the high excitation energy of
the compound nucleus resonance results in fluctuations of shape and
radius of the nuclear potential, which lead to a width
of the eigenvalue of the halo isomer comparable to the compound nucleus
resonance \cite{segre77}. The direct overlap results in strong coupling
of the two states.

While the usual width is typically $\approx$100 meV, which
corresponds to a lifetime of 10 fs, we expect for a halo isomer
bound by 1 eV a lifetime of about 10 $\mu$s and for an isomer with
1 keV binding energy a lifetime of about 300 ps, due to the much
reduced probability in the range of the nuclear core.

Let us first make a very optimistic assumption: We find a neutron
halo isomer with a binding energy $-S_n$= 1 eV. Such an isomer
would have a wave function extending out to $\approx 4000$ fm due
to tunneling. The originally populated compound nuclear resonance,
acting as a doorway state, with a typical radius of 6 fm, would
expand to the neutron halo state only with a very weak overlap.
While the compound states have typical $\Gamma_{\gamma}$ widths of
100 meV \cite{segre77}, the halo state will have a $\sim 10^9$
times longer half life of about 10 $\mu$s. The lifetimes of such
neutron halo isomers are always shorter than the $\sim$~900 s
lifetime of the free neutron, because the weak binding does not
influence the elementary neutron decay. After populating the
isomer it will lose the photon recoil momentum of (6-8) MeV/c by
nuclear stopping \cite{ziegler84} and come to rest. Since these
halo isomers are still smaller than the distance between atoms in
the solid, they should not be destroyed during the stopping
process, where the Coulomb forces keep the nuclei apart.

So far we assumed to find a neutron halo isomer with a binding energy of 1 eV.
However, we may be less lucky and only find an isomer with a larger
binding energy of 1 keV. Looking at the many possibilities for the right
spin combinations in the many nuclei with
A=140-180 the chances for a 1 keV isomer will be rather high.
In this case, the extent of the neutron wave function will be
a factor of 30 smaller and has a typical radius of $\sim$100 fm.
Furthermore, the isomeric lifetime will be reduced by a factor
of $\sim(30)^3$ to about 1 ns. Everything stays the same, except that the
second photon beam would not be a laser but a small second Compton
backscattering X-ray beam. Even a laser-driven, relativistic electron mirror
with coherent Compton back-scattering may be used \cite{habs08}.

\begin{table*}[t!]
\caption{Table of halo Isomers with stable ground state for A=160-180.
The column ``\%'' lists the abundance of the particular isotope in \%,
$I_g$ is the ground state spin, $I_{core}$ the spin of the (A-1) core with
the known excitation energy $E_{core}$ of the excited core,
$S_n$ are the tabulated neutron binding energies from Ref.\cite{audi03}.}
\bigskip
\begin{center}
\begin{tabular}{ccccccccccc} \hline
A  &Z & \%   & $I_g$ &$I_{core}$&$I_{core-g}$&$E_{core}$ & $S_n$ &$E_{tot}$&$I_{i}$&$I_{i0}$ \\
  &  &      &       &        &               & [keV]     &[keV]  & [keV]&&\\ \hline
160&Dy&   2.3&$0^+$  &$3/2^-$&$3/2^-$  &    0.00   &8575.9 & 8575.9 &$1^-$  &$1^-$    \\
161&Dy&  19.0&$5/2^+$&$1^-$     &$0^+$       &1285.60    &6454.39&7739.99  &$3/2^-$&$1/2^+$  \\
161&Dy&  19.0&$5/2^+$&$2^-$     &$0^+$       &1264.73    &6454.39&7719.12  &$5/2^-$&$1/2^+$  \\
161&Dy&  19.0&$5/2^+$&$3^-$     &$0^+$       &1286.69    &6454.39&7741.08  &$7/2^-$&$1/2^+$  \\
162&Dy&  25.5&$0^+$  &$3/2^-$   &$5/2^+$     &  74.58    &8196.99&8271.54  &$1^-$  &$2^+$    \\
162&Dy&  25.5&$0^+$  &$1/2^-$   &$5/2^+$     & 366.95    &8196.99&8563.94  &$1^-$  &$2^+$    \\
163&Dy&  24.9&$5/2^-$&$2^+$     &$0^+$       &  80.66    &6271.01&6352.67  &$3/2^+$&$1/2^+$  \\
164&Dy&  28.1&$0^+$  &$1/2^-$   &$5/2^-$     & 351.15    &7658.11&8009.25  &$1^-$  &$1^-$    \\
164&Er&   1.6&$0^+$  &$1/2^-$   &$5/2^-$     & 345.60    &8847.00&9192.60  &$1^-$  &$1^-$    \\
165&Ho& 100.0&$7/2^-$&$4^+$     &$1^+$       & 166.1     &7988.8 &8154.9   &$5/2^-$&$1/2^-$ \\
166&Er&  33.4&$0^+$  &$3/2^-$   &$5/2^-$     & 242.94    &8474.6 &8713.5   &$1^-$  &$2^-$ \\
167&Er&  22.9&$7/2^+$&$4^-$     &$0^+$       &1596.24    &6436.45&8032.69  &$5/2^-$&$3/2^+$ \\
168&Er&  27.1&$0^+$  &$1/2^-$   &$7/2^+$     & 207.80    &7771.32&7979.12  &$1^-$  &$4^-$ \\
169&Tm& 100.0&$1/2^+$&$1^-$     &$3^+$       &   3.      &8033.6 &8036.6   &$1/2-$ &$5/2^+$ \\
170&Er&  14.9&$0^+$  &$1/2^-$   &$1/2^-$     &   0.0     &7257.2 &7257.2   &$1^-$  &$1^-$ \\
170&Yb&   3.1&$0^+$  &$3/2^-$   &$7/2^+$     & 659.63    &8470.  &9130.    &$1^-$  &$3^=$ \\
171&Yb&  14.4&$1/2^+$&$1^-$     &$0^+$       & 1364.50   &6614.5 &7979.    &$1/2^-$&$1/2^+$ \\
172&Yb&  21.9&$0^+$  &$1/2^-$   &$1/2^-$     &   0.0     &8019.46&8019.46  &$1^-$  &$1^-$ \\
173&Yb&  16.2&$5/2^-$&$2^+$     &$0^+$       &  78.74    &6367.3 &6446.0   &$3/2^+$&$1/2^+$ \\
174&Yb&  31.6&$0^+$  &$1/2^-$   &$5/2^-$     & 398.9     &7464.63&7869.5   &$1^-$  &$3^-$ \\
175&Lu&  97.4&$7/2^+$&$3^-$     &$1^-$       & 111.75    &7666.7 &7778.5   &$5/2^-$&$1/2^-$ \\
176&Yb&  12.6&$0^+$  &$1/2^-$   &$7/2-$      & 370.89    &6864.8 &7235.7   &$1^-$  &$4^-$ \\
176&Hf&   5.2&$0^+$  &$1/2^-$   &$5/2^-$     & 125.93    &8165.0 &8290.9   &$1^-$  &$3^-$ \\
177&Hf&  18.6&$7/2^-$&$4^+$     &$0^+$       & 290.18    &6383.4 &8185.4   &$1^-$  &$4^-$ \\
178&Hf&  21.1&$0^+$  &$1/2^-$   &$7/2^-$     & 559.4     &7625.96&8185.4   &$1^-$  &$4^-$\\
179&Hf&  13.7&$9/2^+$&$4^-$     &$0^+$       & 1747.10   &6098.99&7846.09  &$7/2^-$&$1/2^+$ \\
180&Hf&  35.2&$0^+$  &$1/2^-$   &$9/2^+$     & 375.04    &7387.78&7766.82  &$1^-$  &$5^+$ \\
180& W&   0.1&$0^+$  &$1/2^-$   &$7/2^-$     & 221.93    &8412.  &8634.0   &$1^-$  &$4^-$ \\    \hline
\end{tabular}
\end{center}
\end{table*}

 The core nucleus
frequently requires an excited state, to be reached from the
ground state with an $E1$ excitation to the halo isomer, which is
made by coupling the core spin and parity with the $1/2^+$ halo
state to the $E1$ excitation. In a second step, the core excited
state will deexcite to the ground state, leading to the final halo
isomer, where the $4s_{1/2}$ state is coupled to the core ground
state. In Table 2 and Table 3 we have compiled expected halo
isomers. We give the excitation energy ($E_{tot}$, $S_n$) for the
situation with an excited nuclear core and for the core ground
state, assuming that the $4s_{1/2}$ neutron has zero separation
energy. For most nuclei the isomer may have a higher separation
energy and correspondingly a lower excitation energy. For other
nuclei the neutron may be unbound and no halo isomer exists. The
neutron binding energies were taken from the compilation of Ref.
\cite{audi03}. We have also listed the spins and parities of the
halo isomers for the case with an excited nuclear core and the
case after the deexcitation of the core. The lifetime of the core
excited state will be comparable to the halo isomer lifetime.

\subsection{Photo Excitation of Neutron Halo States}

Next we estimate the cross sections for the photonuclear reaction
and the width $\Gamma_{\gamma}$. The cross section for a compound
nucleus resonance populated by photoexcitation
at the resonance energy $E_r$ in the region of
the neutron separation energy $S_n$ is given by the Breit-Wigner
formula \cite{segre77}

\begin{equation}
   \sigma (E_{\gamma})= (\lambda_{\gamma}^2/4\pi)\cdot g \cdot
   \frac{\Gamma_{\gamma}\Gamma_2}{(E_{\gamma}-E_r)^2 +
   (\Gamma)^2/4},
\end{equation}

where $g=\frac{2I_a+1}{2I_b+1}$ is a spin factor for the spin of
the target and the beam and $\lambda_{\gamma}=(\hbar\cdot
c)/E_{\gamma}$ represents the wavelength of the $\gamma$ rays with
energy $E_{\gamma}$. The resonance is excited with the width
$\Gamma_{\gamma}$ and decays by the width $\Gamma_2$. The
resonance has the total width $\Gamma=\Gamma_{\gamma}+\Gamma_2$.

The $\Gamma_{\gamma}$ width has been studied systematically
as a function of $A$ at the neutron separation energy \cite{segre77} and
we obtain $\Gamma_{\gamma}\approx$ 100 meV for nuclei with $A=180$.
The E1 hindrance factor of $\approx 10^{-5}$ is within a reasonable range
but may fluctuate between different levels. Frequently the maximum cross
section times the width is reported, which in our case 
for compound states is about 10 b$\cdot$eV. However, a halo state with
a typical single particle E1 strength has a much larger E1 excitation width
of typically 1 keV and an integrated cross section of $10^5$b$\cdot$eV. 
Thus even if the halo state fractionates into several components the 
strong components will be populated strongly.
The requirement of the halo state to have a rather small binding energy 
limits the number of states into which it fractionates.
The energy spacing of the compound nucleus resonances for a given spin
and parity at the neutron binding energy for A=180 is about D$\approx$ 1 eV
\cite{segre77}.

For about 60 different isotopes we have a chance of populating a halo isomer. 
Now the challenging question will be: what is the lowest binding
energy and thus the best halo isomer? Does it have a separation energy
$S_N$ of 1 eV or 1 keV?

The Doppler broadening of a $\gamma$ transition at room temperature
$kT=1/40$ eV for a nucleus with mass number $A=200$ and  a $\gamma$ energy
$E_{\gamma}$= 8 MeV is

\begin{equation}
\Delta E_{\gamma}=E_{\gamma}\sqrt{(2kT)/m_pc^2A}\approx \mbox{4
eV}
\end{equation}

Thus the line is broadened with respect to the $\gamma$ beam
by a factor of $\approx 40$. Cooling the target to low temperatures
is advantageous, because then we obtain much larger maximum resonance
cross sections and can use correspondingly thinner targets.
On the other hand, the broadening of the
second low-energy photon beam is negligible. Thus we can produce rather
monoenergetic neutrons in the second excitation.


\subsection{The Photodissociation of Neutron Halo Isomers}

The photodissociation of the halo nucleus is described by the
same formulas as for the deuteron \cite{segre77}

\begin{equation}
\sigma (E_{ph})=\left(\frac{8\pi}{3}\right) \left(\frac{((Z\cdot
e)^2\hbar c}{m_n c^2 S_n}\right)
\left(\frac{S_n(E_{ph}-S_n)}{E_{ph}^2}\right)^{3/2}.
\end{equation}

Due to the change from an s- to a p-wave function, we find an energy
dependence for the cross section of (excess energy above threshold)$^{3/2}$
from the phase space factor. The p-wave neutrons are emitted into a
$\sin{\theta}^2$ cone. Due to the small separation energy $S_n$, and the large
charge Z, the cross section is strongly enlarged compared to the maximum
cross section of the deuteron photodissociation of 2 mb.
For $E_{ph}=2 S_n$ we reach this maximum cross section
$\sigma_{max}=\frac{\pi}{3}(\frac{((Z\cdot e)^2\hbar c}{m_n c^2 S_n})$.
For the large electric field of the coherent photon beam a much more
polarized configuration with a more peaked neutron distribution is obtained.


\subsection{Detection of Neutron Halo Isomers}

For a first identification of halo isomers, we want to
measure the delayed $\gamma$ decay of the isomers and their lifetimes.
Other isomers at lower excitation energies may be used
for lifetime calibration. If in the future $\gamma$-beams with
smaller band width become available, these halo isomers can
be populated more selectively.
While the prompt $\gamma$-decay of the polarized target( due to the
excitation with the polarized $\gamma$-beam)
shows a pronounced angular distribution with minima for the very
short-lived compound nuclear resonances, the polarization
of the halo isomers, due to their lifetime, will be partly lost, allowing
a detection with less background in the minima of the prompt $\gamma$ angular
distributions. Thus the $\gamma$-decay measurements should allow for a
convincing identification of possible halo isomers.

 Experimentally, we first want to identify the new class of neutron halo
isomers by their unique delayed $\gamma$-cascade. Once we have
identified some of these isomers, a more systematic search guided
by a matched theory will become possible and we will be able to explore
their properties.

In a later second step, the release of the weakly bound neutrons could
be studied using a $\approx 0.5$ keV X-ray beam, produced in parallel
from the electron bunches with an undulator. Here the neutron emission
in the direction of the E-field of the undulator radiation, the neutron
spectrum and angular distribution could be measured, leading to a first
realization of this principle for a new neutron beam.

\section{Properties of the New Neutron Beam}

Let us now roughly estimate the flux and brilliance of this
micro-neutron beam: While we start with $10^{13}\gamma$-quanta/s
of about 7 MeV and a band width of about 7 keV ($\Delta E/E=10^{-3}$), 
we excite compound nucleus resonances with a width of $\sim$ 100 meV and thus
should end up with an intensity of $10^8$ isomers/s in a target
spot of $\sim$0.1 mm diameter. The p-wave neutrons are emitted
with (100 mrad)$^2$ opening angle and a band width better than
0.1\%. Thus we estimate a rough average brilliance of $\sim
10^6$/[(mm mrad)$^2$ 0.1\% BW s]. We reduce this value by one order
of magnitude to $\sim 10^5$/[(mm mrad)$^2$ 0.1\% BW s], because we
may not always reach 0.1\% BW. This is about 2 orders of magnitude
better than the best average brilliance of a reactor, e.g. the
H12 beam line of the ILL reactor \cite{ILL88}. Due to the target
isomer thickness of $\sim$1 mm (required to absorb all resonant
photons of the compound nuclear resonance) and the slow movement
of the neutron with about 2000~m/s, the monoenergetic neutron beam
will be pulsed with 1$\mu$s pulse duration. Thus the peak
brilliance will be further increased by a factor of $10^6$
compared to the average brilliance reaching a peak brilliance of
$\sim 10^{11}$/[(mm mrad)$^2$ 0.1\% BW s] (see Fig.~\ref{fig1a}).

For a 1 keV halo isomer it would be more difficult to realize a
second $\gamma$-beam with an energy of 1 keV, which strikes the
target 1 ns after the first $\gamma$-beam of about 7 MeV for the
production of the neutron halo isomers. In principle, such a second
X-ray beam could be realized again with a band width of $10^{-3}$.
Thus we would obtain a neutron beam facility with the same flux
and brilliance, however requiring not a laser but a second X-ray beam.

If such a neutron beam facility would be realized, many parameters
could be further optimized. The neutron pulse duration of $\sim 1
\mu$s was determined by the target thickness. One could envision a
stack of many much thinner neutron halo isomer layers, providing a
macro-bunch of neutrons with micro-bunches of much shorter duration, ns instead
of $\mu$s. However, then also the target thickness has to be very small (nm), 
to avoid a larger delay and spread of the neutron pulse when
crossing the target.

The intense, brilliant second photon beam together with the large halo cross
section can be used to transfer part of its high brilliance to
the neutron beam. A suitable chirp of this laser can
improve the band width of the neutron beam,
making also use of the energy dependence of the photodissociation cross
section. Also different effective target thicknesses with respect to the
$\gamma$-beam and the second photon beam can improve the brilliance by tilting
the target.

If we find neutron isomers with a very low
 binding energy, extremely monoenergetic or extremely
cold neutron beams could be launched with monoenergetic laser pulses
of longer duration.

For a coherent photon field of the second beam a much more
directed neutron emission in the laser polarization direction may
occur due to the collective alignment of the charged core nucleus
with respect to the halo neutron and due to the large dipole
moment, resulting in $\approx 10^4$ times higher brilliances. By
the direction of the $E$-field of the second photon beam,
releasing the neutrons, we can select into which neutron guide the
neutrons are injected. For the flux and brilliance estimates of
the neutron beam, we used a $\gamma$-intensity of
$10^{13}\gamma$-quanta/s, while with the 100 mA ERL a 500 times higher
flux and a 50 times smaller band width come into reach, increasing
the neutron yield further by $\approx 10^4$. Thus $\approx 10^8$
times better neutron beams compared to the conservative estimates
of Fig.~\ref{fig1a} may be reached in 10 years from now (see
Fig.~\ref{fig4} and Fig.~\ref{fig5}).

It is easy to obtain without intensity loss fully polarized neutron beams.
If we Compton back-scatter a fully polarized laser beam from the relativistic 
electron bunches, we obtain fully polarised $\gamma$-beams. If we excite
from a $0^+$ nuclear groundstate the fully polarised $1^-$ halo isomers, we 
obtain fully polarized neutrons, because of their orbital angular momentum l=0.
By switching the polarisation of the primary laser beam, we can switch
from shot to shot the polarisation of the neutron beam. Since the spin of the
neutron is coupled to its magnetic dipole moment the detailed manipulation
of the  polarisation of the neutron beam becomes available.

\section{Physics with the New Brilliant, Pulsed Micro-Neutron Beam}

In neutron scattering there is a large development towards cold
and thermal micro-neutron beams for studying the structure and dynamics 
(excitations) of small samples under extreme conditions, for example in
the area of solid-state and soft-matter physics. Here in
particular the large field of reflectometry, small-angle neutron
scattering (SANS) and diffraction will profit from highly
brilliant and small beams. A large research field in fundamental
physics as well as in applied physics can be opened up by such a
new facility with a large user community with long-term experience
in the scattering of cold and thermal neutrons.

The reasons why the interest in brilliant neutron beams with a
diameter of the order of $30\ \mu{\rm m} < d < 1$ mm has risen
recently rather dramatically are manifold:
\begin{enumerate}
\item[i)] Modern materials exhibiting functional properties,
      showing for example a giant or even colossal magneto-resistance
      or magneto-electric coupling can often only be grown in small
      quantities \cite{canfield2009}. For the investigation of the 
      magnetic and lattice
      dynamical properties, however, neutron scattering is the most
      suitable technique.
\item[ii)] In the field of quantum phase transitions or in geoscience,
         samples are exposed to pressures as high as 30 MPa.
        Applying high pressures involves small samples. To distinguish the
        signals from the sample and the pressure cell, an excellent focusing
        of the neutron beam is required \cite{niklowitz2009}. Of course, a high
        intensity is mandatory.
\item[iii)] Biological samples and multilayers are of major
        interest. Obviously, these samples can only be prepared
        in small quantities.
\item[iv)] A major limitation for the spatial resolution in neutron
         imaging is imposed by the coarse resolution of the
         detectors \cite{radiography05}. Using micron-sized beams
         with a large divergence would allow to apply a cone geometry
         as used in medicine \cite{boeni08,kardjilov10}. The magnification 
         of this setup helps
         to overcome the limited resolution of the detector.
\end{enumerate}

Unfortunately, the presently available neutron beams have a low
brilliance (Fig.~\ref{fig1a}), i.e. they have large cross
section of typically 50 mm$\times 150$ mm and  low divergence.
When adapting the beam size to a typical sample size of the order
of 10 mm$\times 10$ mm using focusing guides, the flux can be
increased, however, due to Liouville's theorem, the divergence is
increased as well~\cite{boeni04,boeni08}. Therefore, experiments
requiring an excellent momentum resolution like SANS and
reflectometry cannot profit much from focusing.

It is exactly the high brilliance of the proposed $\gamma$-beam
facility for the production of neutrons that may lead to major
improvements in future neutron scattering techniques similarly as
it occurred with the invention of synchrotron sources in the field
of x-ray scattering.

\begin{figure}[b!]
\centerline{\includegraphics[width=0.5\textwidth]
{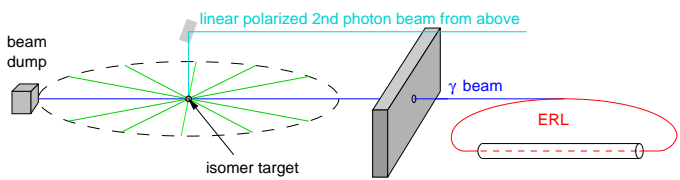}}
   \caption{3D picture showing how the second photon beam releases
  the neutrons from the neutron halo isomer target. The
  polarization of the photon beam injects the neutrons into the selected
  neutron guide.}
   \label{fig9}
\end{figure}

\begin{figure*}[t!]
\centerline{\includegraphics[width=0.95\textwidth]{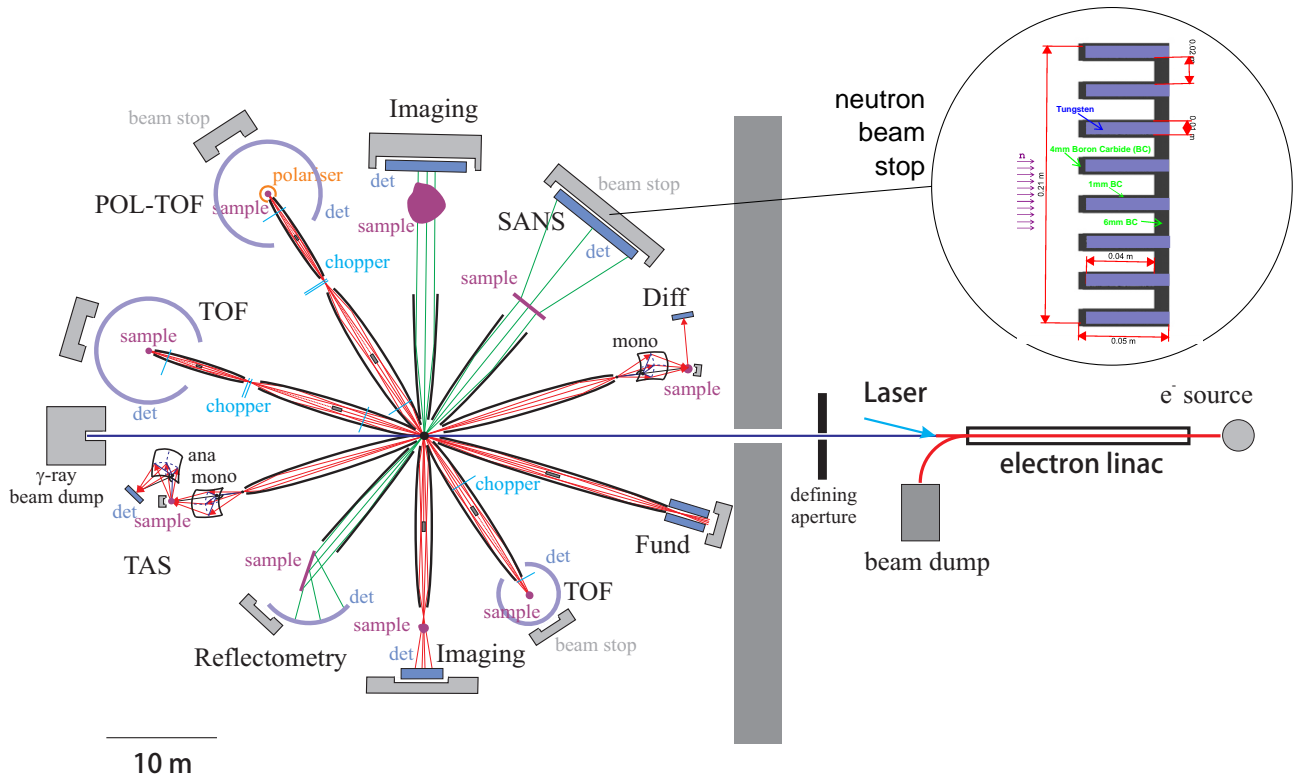}}
   \caption{Experimental setup of a brilliant pulsed micro-neutron
      beam facility. The intense electron beam is shown schematically in
      red. The brilliant $\gamma$-ray beam shown in dark blue hits the neutron
      converter target releasing photo neutrons. A possible arrangement of
      beam lines for neutron science are exemplarily shown 
     (TOF = Time of Flight,
      TAS = Triple Axis Spectrometer, SANS = Small Angle Neutron
      Scattering camera, Diff = Diffractometer, Fund = Fundamental or nuclear
      physics). Beams with low and high divergence are extracted with parabolic
      (green beams) or elliptic (red beams) guides, respectively.}
   \label{fig10}
\end{figure*}

As described above, the second photon beam releases the neutrons
from the isomer target into a very small solid angle depending on
the polarization direction of the second photon beam
(Fig.~\ref{fig9}). The neutrons will feed elliptic or parabolic
neutron guides that transport the neutrons to the various beam
lines for neutron scattering and imaging (Fig.~\ref{fig10}). 
These neutron guides can tolerate a high neutron flux and do not 
transport higher energy neutrons. Since
the E-field of the second photon beam oscillates into opposite
directions, we always feed to oppositely directed neutron guides at
the same time, which should be designed for neutron beams with
similar timing and energy structure. Choppers may be used to
further tailor the properties of the neutron beams. At the end of 
each neutron beam we place an efficient beam dump for thermal neutrons.
A typical beam dump is shown in the insert of Fig.~\ref{fig10},
where the neutrons are captured in boron carbide, while the low energy
capture $\gamma$ rays (0.5 MeV) are absorbed in tungsten. 
Thus even at high neutron intensities an operation without
complex radiation safety licensing procedures seems possible.

Basically, for parabolic or elliptic guides the transport of the
neutrons involves only one reflection in contrast to straight
guides, thus minimizing the reflection losses. These guides extract
neutrons very effectively, because i) only one reflection of
neutrons is involved and ii) the neutron target is located in the
focal point of the guide system. The second focal point of the
elliptic guides will be either the position of a chopper, the
position of the sample, the virtual source of a focusing
monochromator or the source for conical imaging \cite{boeni08}.

The direct line of sight to the target may be interrupted by means
of a beam stop inside the guide. Thus all other radiation
originating from the converter target, like $\gamma$ quanta or
other neutrons are strongly suppressed. Moreover, the smearing of
the time-structure of the ultra-sharp neutron beams is reduced,
because the flight path of the neutrons does not depend on the
divergence. For small-angle neutron scattering or reflectometry, a
parabolic guide geometry is more appropriate to expand the
brilliant micro-beam to a larger size and to decrease its
divergence. Actually, by appropriate design of the guides, the
phase space of the neutrons can be adapted to any beam size
ranging from the size of the production target $d \simeq 100\
\mu$m to millimeters. Using an advanced Kirkpatrick-Baez geometry
for the guides \cite{ice2009}, the beams may be effectively
focused further down to a few $10\ \mu$m or less.

For an efficient extraction of the neutrons, the guides have to
start as close as possible to the converter target. Here, the
radiation load is very high and may lead to the destruction of the
glass that is presently used as substrate for the neutron guides.
Recently it became possible to superpolish metal substrates of
aluminum or iron to an atomic level \cite{schanzer2010} allowing
to start with the guides very close to the target. To further
reduce the background from the converter target, one may tailor the
guides such that only a limited wavelength band of neutrons is
transmitted. This is achieved by using bandpass supermirrors, i.e.
coatings with a laterally and perpendicularly changing $d$-spacing
of the multilayer structure \cite{schneider2009}.

In order to compare the performance of the new source with a high-flux 
reactor, we assume that the converter target ($d \simeq 0.1$
mm) produces a neutron beam with a time averaged brilliance of
$\sim 10^5$ neutrons/[(mm mrad)$^2$ 0.1\% BW s] being the source
for an elliptic guide system feeding a neutron scattering
experiment. The solid angle of the emitted neutrons is
approximately $S_{CT} = 10^4$ mrad$^2$. We consider a beam line
using neutrons with an energy of 81 meV ($\lambda = 1$ \AA).
Assuming supermirrors with index $m = 7$ \cite{snag2010}, the
angle of total reflection of the supermirror is $0.7^0$. An
elliptic guide accepts neutrons with a divergence of at least
$\simeq 2^0$, corresponding to a solid angle $S_A = 1200$ mrad$^2
\ll S_{CT}$. If a beam with a bandwidth of 2 meV is refocused on a
spot with a diameter $d = 0.1$ mm, we obtain a flux of $3\cdot
10^9$ /[(mm$^2$)s] and a divergence of $2^0$. This is indeed an
exceptionally high flux. For a beam size $d \simeq 10$ mm a similar 
estimate yields a flux of $3\cdot 10^5$/[(mm$^2$)s] and a very
small divergence of $0.02^0$. The upgraded triple axis
spectrometer IN8 at the ILL provides a flux at the sample position
of $F_{IN8} = 6.5\cdot 10^{6}$/[(mm$^2$)s] for $\lambda = 1.53$
\AA, however, with a very large divergence \cite{IN8} and poor
energy resolution. Therefore, the converter target may be
competitive with continuous sources even for the investigation of
large samples. The yield for pulsed neutron beam operation will be
even higher: According to Fig.~\ref{fig1a}, the converter target
will deliver orders of magnitude higher intensities than
spallation sources even for large beams.

Next we compare the performance of the converter target for small
angle neutron scattering. Here, beams with a small divergence are
essential. Assuming a brilliance of $\simeq 10^2$ neutrons/[(mm
mrad)$^2$ 0.1\% BW s] at $\lambda = 5$ \AA\ (Fig.~\ref{fig1a}a), we
end up with a flux of $3\cdot10^5$/[(mm$^2$)s] for a beam size of
1 mm, a bandwidth $\Delta \lambda/\lambda = 10\%$, and a
divergence of $0.2^0$. For a sample size of 10 mm, the flux
decreases to $3\cdot10^3$/[(mm$^2$)s], however, with an excellent
divergence as small as $0.02^0$. For comparison, the flux at the
SANS-instrument KWS-1 at the FRM II is quoted to be $10^3$/[(mm$^2$)s]
$< F < 10^5$ /[(mm$^2$)s] depending on the resolution \cite{KWS1}.
Clearly, the flux of SANS at a converter target is comparable to
equivalent beam lines at high-flux reactors. However, we have an
intrinsically pulsed beam and in 10 years from now $10^{8}$ times
better values might be achieved.

The new type of neutron source can be operated in a pulsed mode as
well as in a pseudo CW-mode. The former technique provides an
overview of $S(\bf {Q},\omega)$ in a large area of $\bf {Q}$ and
$\omega$-space, while the latter method is usually more appropriate
for the detailed investigation of $S(\bf {Q},\omega)$ at
particular positions in reciprocal space. Because the pulse
structure and the repetition frequency of the electron and photon
sources involved can be easily adjusted in a wide range of
parameters, the operation of the converter target can be well
adapted to the requirements of the various beam lines. By
switching the polarization of the $\bf E$-field of the laser
during operation, the various beam lines may be fed with neutrons
continuously, however, at a lower pulse rate that might reduce the
problem of frame overlap. Last but not least, the possibility to
fully polarize the neutrons during their production with polarized
$\gamma$-rays without loss of brilliance allows to study magnetic
materials and soft materials with strong incoherent scattering. By
analyzing the polarization of the scattered neutrons, information
on the detailed orientation of the magnetic moments in the sample
and their interaction with the lattice degrees of freedom can be
studied.

\section{Conclusions}

The new neutron facility opens a broad range of new possibilities
for neutron physics. Once successful, neutron sources could be
built even at universities resulting in a large scientific
community. If one compares the investment and running costs of the
proposed neutron facility with the costs of reactors or spallation
sources, they will be $\approx$ 2 orders of magnitude smaller.
Moreover, it is also tremendously cheaper than any proposed
neutron source based on inertial fusion \cite{taylor2007}. The big
advantage of the new facility is that it produces only small
amounts of radioactivity and radioactive waste and thus requires
only small efforts for radioprotection, i.e., very different from
present reactor or spallation facilities. The new source can be
operated as a multi-user neutron facility and could be several
orders of magnitude more brilliant than the best existing neutron
sources.

The same electron linac and recycling loop could drive in parallel
a brilliant X-ray and positron source, where a 2-3 MeV
$\gamma$-beam via the $(\gamma,e^+e^-)$ reaction produces
positrons in a converter target. Such a positron source would 
be similar to the NEPOMUC positron source at the 
FRM II~\cite{hugenschmidta08,hugenschmidt08},
however, the $\gamma$-rays would originate from a brilliant $\gamma$-beam 
and not from neutron capture. This allows to address the same 
targets with very different beams and analyzing techniques and makes neutron 
sources using halo isomers a unique possibility. In particular, these new
sources have the potential, similarly as 30 years ago the
synchrotron sources for X-ray scattering, to increase the flux of
neutron beams by orders of magnitude in the next decades to come.
Already with existing technologies, a neutron scattering facility
based on halo isomers may deliver more intense beams for the
investigation of small samples than today's most powerful neutron
sources.

\vspace{10mm}

{\bf Acknowledgement}\\

We acknowledge helpful discussions with C.~Barty, R.~Hajima and
K.~Schreckenbach. We enjoyed the close collaboration with V.~Zamfir,
who is heading the ELI-NP project.
We were supported by the DFG Clusters of Excellence:
Munich Centre for Advanced Photonics (MAP) and UNIVERSE.

\end{document}